\documentclass[proceedings]{JHEP}
\usepackage{epsfig}
\conference{TMR9: Quantum aspects of gauge theories, 
supersymmetry and unification}
\title{The Holographic RG flow to conformal and non-conformal theory\footnote{Talks presented at the TMR conference in Paris, September 99.}}
\author{M. Petrini and A. Zaffaroni\\
M.P.: Theoretical Physics Group,\\
Imperial College, London SW7 2BZ, UK.\\
A.Z.: INFN, Sezione di Milano,\\via Celoria 16,20133, Milano, Italy.\\
}
\abstract{We review some aspects of the AdS
supergravity description of RG flows. The case of a flow to an IR 
CFT can be rigorously studied within the framework of supergravity.
Here we discuss various central charges of the conformal theory 
(included the usually neglected ones) and we
compare them with QFT expectations. The case of
flows to non-conformal theories is more problematic in that one usually
encounters a naked singularity.
We mainly focus on the flow
to an IR N=1 super Yang-Mills theory. We discuss the properties
of the solution and we briefly comment on the fate of the singularity.
We also compare the supergravity results with the expectations of
an N=1 SYM at strong coupling.}
\begin{document}
\newcommand{\Tr}{\mbox{Tr\,}}
\newcommand{\beq}{\begin{equation}}
\newcommand{\eeq}[1]{\label{#1}\end{equation}}
\newcommand{\bea}{\begin{eqnarray}}
\newcommand{\eea}[1]{\label{#1}\end{eqnarray}}
\renewcommand{\Re}{\mbox{Re}\,}
\renewcommand{\Im}{\mbox{Im}\,}
\section{Introduction}
The AdS/CFT correspondence has deserved some surprises
when extended outside the realm of stric-\
tly conformally invariant theories.
The study of the supergravity dual of RG flows has flourished,
both in the concrete application to SYM theories 
and in a general setting \cite{gppz1}-\cite{verlinde}.
Asymptotically AdS$_{d+1}$ backgrounds, breaking the full $O(d,2)$ invariance
but preserving at least $d$-dimensional Poincar\'e invariance,
describe RG flows for a $d$-dimensional CFT.
These supergravity solutions with an asymptotic  AdS region 
have a double QFT interpretation: deformations of an UV fixed point versus 
the same theory in a different vacuum \cite{bala,kw2}. Both cases
have been extensively studied. Many results have
been obtained upon reduction to a $d+1$-dimensional effective theory,
where the RG flow can be studied in terms of a theory of scalar fields
coupled to gravity. In this simple set-up, the RG flows are 
identified as domain-walls interpolating
between AdS$_{d+1}$ vacua (or approaching infinity on one side), and
general results are very easy to obtain. 
The correspondence defines a {\it holographic}
scheme, where beta and $c$-functions have a natural definition. 
A $c$-theorem, for example, can be easily proven \cite{gppz1,freed1}.
Moreover, it is possible to obtain the quantum field theory RG equations 
from supergravity \cite{verlinde}\footnote{Notice that the {\it holographic} 
beta and $c$-functions do not need to coincide with analogous 
functions defined in schemes that are more natural from the QFT point
of view \cite{ans}.}.

The study of RG flows between CFTs (at large $N$ and strong coupling)
can be rigorously performed using supergravity. The phase space of massive
deformations of the N=4 SYM theory has been throughly investigated
and several IR fixed points have been found \cite{gppz1,dz,pilch,lust,freed1}.
The results are on solid grounds because supergravity is valid all along
the RG flow.
Still problematic is the precise mapping of some QFT couplings to
supergravity quantities.
For example, it is still unclear what in supergravity corresponds
to the running of the gauge coupling. 

Most of the unsolved problems concern the flows to non-conformal theories,
where supergravity is invalidated by a (typically naked) singularity in
the IR region of the flow. 
Solutions flowing to infinity for a generic
5$d$-Lagrangian are certainly a dense set in the space of solutions. The
full recipe for selecting the {\it physical} ones is still 
unclear\footnote{A criterion for selecting physical
solutions has been recently proposed in \cite{gubbb}.}.
The distinction between deformations and vacua of an UV fixed point
helps but does not solve the problem. Supersymmetric and 
supersymmetric-inspired solutions however are uniquely selected
because the equations of motion can be reduced to first order ones 
\cite{freed1,townsk}.
N=4 Coulomb branch solutions have been studied in
\cite{freed2,brr,cvet}. Here we focus on the flow to N=1 SYM.
Despite the singularity, we obtain a good qualitative agreement with
quantum field theory expectations already at the level of supergravity.

Since singularities are apparently unavoidable in interesting
supergravity solutions, it is mandatory to understand their fate
in the full string theory, where they must be resolved. Available options
are the chance that the singularity is an artifact of the
dimensional reduction to 5 dimensions, mechanisms such that proposed
in \cite{pjp} and, more generally, some help from string corrections.

The supergravity solutions with an asymptotic AdS region certainly have
many other applications.
Relaxing the $d$-Poincar\'e invariance, we have examples of
RG flow due to finite temperature. This is indeed the firstly proposed
me-\
thod for discussing non-conformal theories from AdS \cite{w2} and
the one not suffering from unpleasant singularities.
Cutting the AdS-boundary, we can describe CFTs coupled to gravity and
make contact with the large extra-dimension scenario \cite{rs}. We will
not discuss this issue here, but we simply notice that singular solutions
have been recently considered in this context.

\section{RG Flow from 5$d$ Supergravity}
In general, we interpret the $(d+1)$-th coordinate $y$ of AdS$_{d+1}$ as 
an energy scale \cite{malda,PP}. RG flows between CFTs then 
correspond to type II
or M-theory supergravity solutions interpolating (along $y$) between 
AdS$_{d+1}\times_W H$ vacua.  

The very first example of RG flow in the AdS/CFT correspondence
is manifest in the multi-centre supergravity solution for  D3-branes
\cite{malda}. This represents the Coulomb branch of N=4 SYM. Given
two sets of $N$ and $M$ branes at different points, the near-horizon
geometry is  AdS$_5$ with radius $\sim\sqrt{N+M}$ far from both sets of
branes, and AdS$_5$ with radius $\sim\sqrt{N}$ near one set.
In QFT this is the RG flow between the $U(N+M)$ N=4 CFT in the UV,
where the Higgs VEVs can be neglected, and the $U(N)$ N=4 CFT in the IR.
A more sophisticated example was found in \cite{kw1}. A supergravity solution
interpolating between AdS$_5\times S^5/Z_2$ and AdS$_5\times T^{1,1}$
was also interpreted on the QFT side as a RG flow between CFTs.
It is a supersymmetric massive deformation
of the N=2 $SU(N)\times SU(N)$ theory corresponding to a $Z_2$ orbifold
of N=4 SYM  which flows to an N=1 IR fixed point. Many successful checks
of this interpretation have been performed \cite{kw1,g,gk,sergio}.

However, interpolating 10$d$ backgrounds are difficult to find.
Sometimes dimensional reduction to 5 dimensions helps.

The RG flow has a natural description in 5$d$. \\
Consider a certain UV CFT and suppose we have the corresponding 
5$d$ Lagrangian  
and that it contains all the fields/modes we are interested in. 
The effective 5$d$ Lagrangian
we need is just the most general Lagrangian for scalars coupled to gravity
\beq
L= \sqrt{-g}\left [ -{R\over 4}+ {1\over 2}g^{IJ}\partial_I\lambda_a \partial_J
\lambda_b G^{ab} + V(\lambda)\right ].
\eeq{lagr}
The scalars  $\lambda_a$ can either be the {\it massless} modes or
Kaluza-Klein modes of the
compactification to 5 dimensions. The form of the potential depends on
the particular case we are considering.  
We may have, for example, N=8 gauged supergravity, which describes
N=4 SYM and most of its bilinear relevant operators (almost all of the
masses for scalars and fermions). 
Or we may have an N=4 theory describing the orbifold
$R^4/Z_2$ and the supersymmetric mass term that drives the theory to an
N=1 IR fixed point. Or else we may have the Lagrangian for some of
the KK modes.
The interactions among the modes in the graviton multiplet 
in 5$d$ can be found using supersymmetry. In particular, for the N=4 SYM
case, the 5$d$ Lagrangian for the {\it massless} modes is uniquely 
fixed by supersymmetry in the form of the N=8 gauged supergravity \cite{warn}. All mass terms
for the scalars and the fermions contained in the KK
spectrum
are associated to modes in the gauged supergravity.
5-dimensional supersymmetric Lagrangians have been discussed 
also for less supersymmetry,
but the uniqueness of N=8 supergravity is lost and interesting modes
are split into various vector, tensor and hyper-multiplets. One needs
some help from QFT intuition in identifying the right potential.
In principle, $V(\lambda)$ can be obtained for all modes
(often with non-trivial effort) by dimensional reduction from 10 dimensions.

If the UV CFT perturbed by a particular 
operator $O_\lambda$ flows in the IR to another CFT, the potential $V$
must have a critical point for non-zero value of the scalar field $\lambda$.
Analogously, the dual of the flow to a non-conformal field theory is
given by the flow from one minimum of the potential to infinity.

\begin{figure}[h]
\epsfxsize = 2.8in
\epsffile{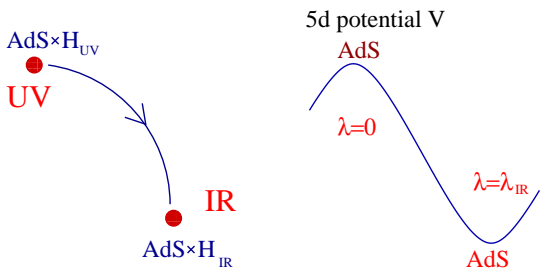}
\caption{Schematic picture of the RG flow.}
\end{figure}


\noindent The 5$d$ description of the RG flow between conformal theories
is a kink solution, which interpolates between the two critical points. 
A 4$d$ Poincar{\'e} invariant metric is
\beq
ds^2= dy^2 + e^{2\phi(y)}dx^\mu d x_\mu , \;\;\;\mu =0,1,2,3.
\eeq{hjk}
AdS corresponds to $\phi =y/R$. 
We then look for solutions with asymptotics: $\phi(y)\rightarrow y/R_{UV,IR}$ 
for $y\rightarrow\pm\infty$; $\lambda(y)\rightarrow 0$ for
$y\rightarrow\infty$, while $\lambda(y)\rightarrow \lambda_{IR}$ for
$y\rightarrow -\infty$. We associate larger
energies with increasing $y$.

The equations of motion for the scalars and the metric read
\bea
 \ddot\lambda_a+4\dot\phi\dot\lambda={\partial V\over\partial\lambda_a},
\nonumber\\
6(\dot\phi )^2=\sum_a(\dot\lambda_a )^2 -2V.
\eea{I7}
With the above boundary conditions
and a reasonable shape of the potential,
a kink interpolating between critical points always exists \cite{gppz1}.

As an example of flows between conformal field theories, we can discuss
 the mass
deformations of N=4 SYM. These  can be studied in the context of N=8
gauged supergravity, where the form of the potential $V$ is known.
N=8 gauged supergravity \cite{warn} is the low energy effective
action for the ``massless'' modes of the compactification of type IIB on
$AdS_5 \times S_5$. It is believed to be a consistent
truncation of type IIB on $S^5$ in the sense that every
solution of the 5$d$ theory can be lifted to a consistent 10$d$ type IIB
solution.
Five-dimensional gauged supergravity has 42 scalars, which
transform under the N=4 YM R-symmetry $SU(4)$ as
$\underline{1},\underline{20},\underline{10}$.
The singlet is associated with the marginal deformation
corresponding to a shift in the coupling constant of the N=4 theory.
The mode in the $\underline{20}$ has mass square $M^2=-4$ and is
associated with a symmetric traceless mass term for the scalars
$\Tr\phi_i\phi_j$, ($i,j=1,...,6$) with $\Delta = 2$.
The $\underline{10}$ has mass square $M^2=-3$ and corresponds to 
the fermion mass term $\Tr\lambda_A\lambda_B$, ($A,B=1,...,4$) of
dimension 3.
Thus the scalar sector of N=8 gauged supergravity is enough to discuss
at least all mass deformations
that have a supergravity description\footnote{ The only missing state is 
$\Tr \sum_i^6\phi^2_i$, the prototype of a stringy states in the 
correspondence. Even without this state, we can study almost all 
massive deformations of the N=4 theory and all these deformations
can be described by just the Lagrangian for the massless multiplet.}.

The scalar potential $V$ in eq.(\ref{lagr}) is
known and it turns out to have only isolated minima (apart from one flat
direction, corresponding to the dilaton). 
Up to now,
all critical points with at least $SU(2)$ symmetry have been classified
\cite{pilch}. There is a central critical point with
$SO(6)$ symmetry and with all the 
scalars $\lambda_a$ vanishing: it corresponds to the unperturbed N=4 YM theory.
There are three N=0 theories with residual symmetry $SU(3)\times
U(1)$, $SO(5)$ and $SU(2)\times U(1)^2$. They correspond to non-zero VEV
for some of the scalars in the $\underline{10}$, $\underline{20}$, and $\underline{10} +\underline{20}$, respectively.
Then there is an N=2 point with symmetry $SU(2)\times U(1)$, obtained
giving VEV to scalars in the $\underline{10} + \underline{20}$
\cite{pilch}. 
According to the AdS/CFT correspondence, 
these other minima should correspond to IR conformal field theories\footnote{The symmetries of field
theories can be read from those of the supergravity minima according to the
correspondence : gauge symmetry in supergravity $\leftrightarrow$ global
symmetry in field theory, supersymmetry in supergravity
$\leftrightarrow$ superconformal symmetry in field theory.}. The following IR 
CFT theories can be obtained as mass
deformations of N=4 SYM:
\begin{itemize}
\item{
Three N=0 theories with symmetry $SU(3)\times U(1)$, $SO(5)$ and
$SU(2)\times U(1)^2$. All these theories are unstable  and correspond
to non-unitary CFTs. A natural question arises: are all the N=0 critical
points unstable?} 
\item{A stable N=1 theory with symmetry $SU(2)$ 
$ \times U(1)$.
It corresponds to the N=4 theory deformed with
a mass for one of the three N=1 chiral superfields. 
Results and supergravity description \cite{lust,freed1} are almost
identi-\
cal to
the $T^{1,1}$ case, which is just a $Z_2$ projection of this example.}
\end{itemize}

\subsection{Central charges}
In a supersymmetric gauge field theory in 4$d$, the trace and R-symmetry
anomaly are given by \cite{afgj1} 
\bea
T_{\mu}^{\mu}&=&{\tilde{\beta} \over 2g^2} F_{\mu\nu}^2 
+ {c \over 16 \pi^2} W_{\mu\nu\rho\sigma}^2 -
    {a \over 16 \pi^2} \tilde{R}_{\mu\nu\rho\sigma}^2 \nonumber\\
&+& {c \over 6 \pi^2} V_{\mu\nu}^2 +  {b \over 32 \pi^2}
B_{\mu\nu}^2, \\
\partial_\mu \sqrt{g} R^\mu &=& -
{\tilde{\beta} \over 3 g^2} F_{\mu\nu}\tilde{F^{\mu\nu}} 
-{a-c \over 24 \pi^2} R_{\mu\nu\rho\sigma}
\tilde{R}^{\mu\nu\rho\sigma}\nonumber\\ 
&+& {5a-3c \over 9\pi^2} V_{\mu\nu} \tilde{V}^{\mu\nu} 
- {b \over 48 \pi^2} B_{\mu\nu}\tilde{B^{\mu\nu}}.\
\eea{le}
Here $W_{\mu\nu\rho\sigma}$ and $R_{\mu\nu\rho\sigma}$ are the Weyl
and curvature tensors for an external metric $g_{\mu\nu}$ that
couples to the energy-momentum tensor $T_{\mu\nu}$. Similarly
$V_{\mu\nu}$ and $B^{\mu\nu}$ are the field strengths of the external
sources $V_{\mu}, B_{\mu}$ that couple to the R-symmetry and flavour
currents, respectively. $F_{\mu\nu}$ is the gauge field
strength and 
$\tilde{\beta}$ is the numerator of the exact beta-function \cite{nsvz}.

The external anomaly coefficients $a$ and $c$ have a
straightforward interpretation in the dual supergravity theory.

$c$ is the central charge of the CFT, and it is associated with the
cosmological constant at the critical points.
From eq.~(\ref{lagr}),
we can see by a simple scaling that, at least at the 
fixed points, where $ds^2=R^2[dy^2+\exp(2y)\sum_i dx_i^2]$,
\beq
\langle T(x) T(0)\rangle = {c\over |x|^8}\,\,\rightarrow c\sim R^3\sim (\Lambda )^{-3/2}.
\eeq{scaling}
This scaling reproduces the known results for $c$ \cite{g,hs}. More
interestingly, one can prove that for the class of field theories that
have a supergravity dual a $c$-theorem exists. Indeed we can exhibit a
$c$-function that is monotonically decreasing along the flow
\cite{gppz1,freed1}. The $c$-function 
\beq 
c(y)\sim (T_{yy})^{-3/2},
\eeq{LL}
is constructed with the $y$ component of the stress-energy tensor
\beq 
T_{yy}=6(\dot\phi )^2=\sum_a(\dot\lambda_a )^2 -2V.
\eeq{JJJJJ}
At the critical points, where $\dot\lambda_a=0$, 
\beq 
c(y)=c_{UV,IR}\sim (-V)_{UV,IR}^{-3/2}\sim\Lambda_{UV,IR}^{-3/2}, 
\eeq{jj}
and using the equations of motion  ($\ddot\phi < 0$) and the boundary
conditions one can easily check that $c(y)$ is monotonic \cite{gppz1,freed1}.

Let us consider $a$. AdS computations~\cite{hs} showed that $a=c$ for
all CFTs that have an AdS dual.

It is then natural to ask what can AdS/CFT correspondence say about
the coefficient $b$ \footnote{These results have been obtained in collaboration
with D. Anselmi and L. Girardello.}.
The coefficient $b$ is related to the two-point function of the flavour
(global) symmetry currents~\cite{afgj1}. 
According to AdS/CFT correspondence the
R-symmetry and flavour currents are associated to the gauge fields of
the SUGRA Lagrangian
\beq
J_{\mu}, R_{\mu} \leftarrow\rightarrow A_{\mu}.
\eeq{}

\noindent One should then be able to read the $b$ (and $a$) coefficient from the
kinetic terms of the corresponding SUGRA modes. The generic 
5$d$-Lagran-\
gian we are interested in has the following structure 
\beq
L= \sqrt{-g}\left [ -{R\over 4} + \Lambda + f F_{\mu\nu}^2 + 
f_R F_{\mu\nu R}^2 
\right ].
\eeq{lagr2}
Here $F_{\mu\nu R}$ and $F_{\mu\nu}$ represent the kinetic terms for the
fields corresponding to the R-symmetry and flavour symmetry currents,
respectively. 
At the critical points (or generically for a metric of the form
(\ref{hjk})), one obtains by scaling
\beq
\langle J(x) J(0)\rangle = {b\over |x|^6}\,\,\rightarrow b\sim f R
\sim f c^{1/3}.
\eeq{scaling2}
A similar behaviour is obtained for
the R-symme-\
try currents. In this case, supersymmetry~\footnote{The
R-symmetry currents are in the same multiplet as the energy-momentum
tensor.}  implies  $b=c$, and
the previous equation can be used as a check of the consistency of the
procedure. 

The values of the coefficients $f$ and $f_R$ depend on the particular
model under consideration. Consider for example the massive
deformations of N=4 SYM, for which we have the dual supergravity
Lagrangian: that of N=8 gauged supergravity. 
In this case, the kinetic term for the gauge fields is expressed in terms
of the vielbein parametrising the scalar manifold~\cite{grw2}. 
To determine $f$ and $f_R$ we have then to evaluate the
contractions of the vielbein and therefore these coefficients depend on
the critical point 
and on the way the UV $SU(4)$ group is broken (for instance, $SU(4)$ $\rightarrow$
$SU(3)$ $\times$ $U(1)_R$, or  $SU(4)$ $\rightarrow$
$SU(2)$ $\times$ $ U(1)_R$, ...).
We now want to compute the charge $b$ for the global non-abelian
symmetry group preserved a-\
long the flow (e.g. $SU(3)$, $SU(2)$, ...).
The computation of the coefficients $f$
can be performed using the results of~\cite{grw2} for most of the
critical points. Alternatively, using the parametrisation in appendix A
of \cite{freed1}, it is easy to convince themselves that 
\beq
f =e^{4 \alpha}.
\eeq{fF}
Here $\alpha$ is the scalar in the $\underline{20}$ of $SU(4)$
corresponding to a mass term for the scalars in N=4 SYM~\cite{freed1}.
The value of the scalar $\alpha$ and $c$ for the various fixed
points can be found in \cite{pilch,freed1,grw2}. One then gets the
following results for the coefficient $b$ \cite{non}: 
\begin{itemize}
\item{N=1 point with symmetry $SU(2)\times U(1)$. ${b_{IR}\over
b_{UV}}= {3 \over 2}$. This is the only case where comparison with field
theory is possible. Consider a set of N=1 chiral superfields $X_i$ in
the representation $R_i$ of the gauge group and in the representation
$T_i$ of the flavour symmetry group. Then, because of supersymmetry, the
following formula holds~\cite{afgj1} 
\beq
b_{UV} - b_{IR} = 3 \sum_{ij}\left(dim R_i \right) \left[\left(r_i -
{2\over 3}\right) T_{i}^{j}T_{j}^{i}\right],
\eeq{bb}
where $r_i$ is IR R-symmetry charge of the field $X_i$ and
$T_{i}^{j}$ are the generators of the flavour 
group in the representation $T_i$. It is straightforward to check that the
supergravity and the field theory computations agree.}
\item{N=0 theories. For the $SU(3)\times U(1)$, $SO(5)$ and $SU(2)\times
U(1)^2$ symmetric points, we have ${b_{IR}\over b_{UV}}= {2\sqrt{2}
\over 3}$, ${b_{IR}\over b_{UV}}= \sqrt{2}$ and ${b_{IR}\over b_{UV}} =
2$, respectively.} 
\end{itemize}

In~\cite{afgj2} it was observed that for several examples of
supersymmetric gauge theory $b$ increa-\
ses going from the UV to the
IR. This was suggestive of possible anti-$b$-theorem.
The same authors however pointed out that for non-supersym-\
metric gauge
theories $b$ has no universal behaviour, and that also a large class of
supersymmetric theories violates the relation $b_{IR}/b_{UV} >1$. Then it
is not possible to state any anti-$b$-theorem in field theory.
It is interesting to see what are the supergravity results. Consider
first the non-supersymmetric cases. For the point $SU(3) \times U(1)$ we have
$b_{IR}/b_{UV}<1$, which violates the anti-$b$-theorem.
The situation is different for the supersymmetric
point $SU(2)\times U(1)$. In this case the coefficient
$b$ increases along the flow. The same analysis
carried on 
for the massive flow to N=1 super Yang-Mills (see section 4) or  for the
Coulomb branch of N=4 SYM~\cite{freed2} seems to
indicate a similar behaviour. 

Notice that the theories that have a supergravity dual represent a very
restricted class of gauge theories. First of all these theories always have
$a=c$, which is in general not the case in field theory. It has been
argued that the requirement $a=c$ simplifies the structure and OPEs of
a CFT, making it most similar to a two dimensional conformal field
theory \cite{ans2}.
Secondly it has
been suggested (see \cite{freed1} and next section) that all these
theories could be characterised by having a pre-potential. It could then
be possible, and interesting to check, whether an anti-$b$-theorem could
hold for this particular class of gauge theories.

The previous results on $b$ could have been obtained from the analysis
of the Chern-Simons terms of the N=8 Lagrangian, which contain all
information about global anomalies \cite{w1,fz}. In particular,
$b$ can be read from the $SU(2)^2 \times U(1)_R$ anomaly coefficient,
which can be extracted from the Chern-Simon terms.
It is easy to check, using the results in \cite{freed1}, 
that the result for $b$ coincides with the previously
obtained one\footnote{It is crucial
to pay attention
to normalisations and the definition of $U(1)_R$, which varies
from UV to IR.}. Notice that the Chern-Simon terms uniquely determine
the form of a supersymmetric gauge supergravity. From the knowledge
of the global anomaly, we should be able to reconstruct the 
entire AdS Lagrangian for {\it massless} modes
for a given supersymmetric CFT fixed point \cite{fz}. 

\subsection{Vacua and deformations}  
We end this section with a brief discussion of a point that will play an
important role in our analysis, namely the fact that supergravity
solutions can represent both deformations of a CFT and different vacua of the
same theory \cite{bala,kw2}. 
The running of coupling constants and parameters along the RG flow
can be induced in the UV theory in two different ways:
by deforming the CFT with a
relevant operator, or by giving a nonzero VEV to some operators. 
The asymptotic UV behaviour discriminates between the two options.
In the asymptotic AdS-region, we just need a linearised analysis.
A scalar fluctuation $\lambda(y)$ in the asymptotically AdS background 
must satisfy
\beq
\ddot\lambda +4\dot\lambda=M^2\lambda,
\eeq{I2}
where the dot means the derivative with respect to $y$.
The previous equation has a solution depending on two arbitrary
parameters
\beq
\lambda(y)=Ae^{-(4-\Delta )y}+Be^{-\Delta y},
\eeq{I3}
where $\Delta$ is the dimension of the operator,
$M^2=\Delta (\Delta -4)$ \cite{w1,gkp}.
We are interested in the case of relevant operators, where $\Delta\le 4$.
From the basic prescription of the AdS/CFT, we associate
solutions behaving as $e^{-(4-\Delta )y}$
with deformations of the N=4 theory
with the operator $O_\lambda$. On the other hand, solutions asymptotic
to $e^{-\Delta y}$ (the subset with $A=0$) are associated with a different
vacuum of the UV theory, where the operator $O_\lambda$ has a non-zero
VEV~\footnote{We are not careful about 
subtleties for particular values of $\Delta$ \cite{kw2}.}\cite{bala,kw2}.

Since in general the UV-IR interpolating solution is not known, 
it is not even obvious
whether a particular solution corresponds to a deformation 
or to a different vacuum.
For many problems,  we may invoke supersymmetry. 
It helps in finding the solution
all along the flow and in  unambiguously identifying the UV behaviour.
In ref.~\cite{freed1,townsk} 
the conditions for a supersymmetric flow were found.
As usual, a solution for which the fermionic shifts
vanish, automatically satisfies the equations of motion. Moreover, this
shortcut reduces the second order equations to first order ones.
For a supersymmetric solution, the potential $V$
can be written in terms of a superpotential $W$ as
\beq
V = \frac{1}{8} \sum_{a=1}^{n} \left|
\frac{\partial W}{\partial \lambda_a} \right|^2
- \frac{1}{3} \left|W \right|^2,
\eeq{I9}
where $W$ is one of the eigenvalues of the tensor $W_{ab}$ defined in
\cite{grw2}. The equations of motion reduce to
\bea
\dot\lambda_a&=&\frac{1}{2} \frac{\partial W}{\partial \lambda_a},\\
\dot\phi&=& - \frac{1}{3} W.
\eea{I10}
It is easy to check that a solution of eq.(\ref{I10}) satisfies also the
second order equations ~(\ref{I7}).

It is quite plausible and generally assumed that all the supergravity flows
connecting fixed points correspond to deformations of the UV fixed point.
\section{Confining Solutions}
Solutions flowing to infinity represent RG flows to non-conformal
theories, which may exist in various phases in the IR.
These kinds of solution are difficult to classify. In many
cases the asymptotic IR behaviour is known, but the entire solution
along the flow can not be found.
Typically, we encounter a singularity somewhere along the flow.
Many solutions exhibit a logarithmic divergence at finite $y_0$ 
 for the scalar fields,
$\lambda_a\sim B_a \log |y-y_0|$, and the metric, $\phi \sim A \log |y-y_0|$.
There are many criteria for studying the IR properties and the phase
of these solutions. One of them, the Wilson loop, will be discussed 
later. The spectrum can be determined also from two-point functions,
where physical bound states appear as poles.
Poles in the two-point function corresponding to a minimally coupled
scalar, for example,
correspond to $F^2$  glueball masses in the field theory. The analysis of the
spectrum can be reduced, as usual in the AdS/CFT correspondence, to
the solution of a Schroedinger problem \cite{w2,oz}. After a change of
variable $y\rightarrow z$ to the conformally flat metric $ds^2=e^{2\phi
(z)}((dz)^2+(dx)^2)$ and a field 
redefinition $\Phi_k (z) =e^{-3\phi (z) /2}\psi (z)$, the
5$d$ equation for a minimally coupled scalar $\Phi (x,y)=e^{-ikx}\Phi_k (y)$
takes the Schroedinger form 
\beq
( -\partial_z^2 + V(z))\psi= E\psi
\eeq{scr}
where $V={3\over 2}\phi^{\prime\prime}+{9\over 4}(\phi^{\prime})^2$.
The eigenvalues $E$  give the poles in the two-point function
and the spectrum.

\begin{figure}[h]
\epsfxsize = 2.8in
\epsffile{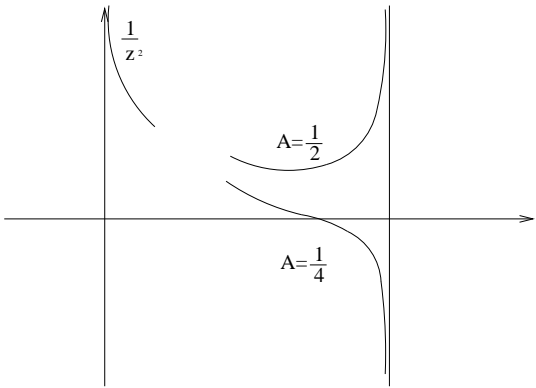}
\caption{The Schroedinger potential in various cases.}
\end{figure}


\noindent The form of $V$ immediately
tells us whether the theory has a  mass gap and a discrete spectrum or 
a continuous one, whether it confines or not. Unfortunately, in very few examples
$V$ is known along the entire flow. We can nevertheless extract some
information from the IR behaviour. For the logarithmically
divergent flows discussed above, if $A<1$, the singularity is mapped to
a finite $z_0$ and we have
\beq
V\sim {3A(5A-2)\over 4(1-A)^2(z-z_0)^2}.
\eeq{vs}
This behaviour looks potentially
dangerous, but,
as discussed in all quantum mechanics textbooks, 
$V\sim k/z^2$ has 
a discrete spectrum bounded from below, provided $k\ge -1/4$.
It is easy to check that, for the logarithmically
divergent flows, this condition is always satisfied.
The value $k=-1/4$ is obtained for
$A=1/4$. This is the value 
that appears in many solutions where the supergravity potential is irrelevant
in the IR \cite{gppz2}, but also in one of the examples of N=4 coulomb branch
in \cite{freed2}. 
If $A>1$, the singularity is mapped to $z=\infty$,
the potential goes to zero and we may expect portions of continuous
spectrum. Clearly, any sensible prediction about the spectrum requires
the full knowledge of $V$. The same Schroedinger equation is to be
considered when looking at generalisations of the RS scenario.
\subsection{Supersymmetric and non-supersymmet-\ ric examples}
We now briefly discuss few examples in the literature. 

In \cite{gppz2}, the class of non-supersymmetric 
solutions where the potential can be
neglected in the IR have been discussed. They all have $A=1/4$.
It was argued that they may exhibit a variety of IR behaviours,
from confinement to screening, depending on the values of
the constants $B_a$. Since we can not follow the solution from UV to
IR, it is difficult to make more meaningful claims. We do not even
know whether these solutions correspond to deformations 
or to different vacua of the UV fixed point.

In the N=4 Coulomb branch solutions discussed in \cite{freed2}, $A$
assumes various values. There is one solution with $A=1/5$, one with $A=1/4$
and all the other have $A>1/4$. The UV behaviour can be unambiguously
determined using the first-order equations~ (\ref{I10}). All these
solutions correspond to different vacua (Coulomb branch)
of the UV fixed point.  

The supersymmetric massive flow from N=4 to N=1 SYM was discussed in 
\cite{gppz3}. It has $A=1/2$. The qualitative properties of the solution
agree with QFT expectations. They are discussed in the
next section.  

Due to the IR singularity, not all the previous solutions are expected to be
{\it physical}. A possible criterion for selecting the {\it physical}
solutions has been proposed in \cite{gubbb}. According to this criterion,
the supergravity potential must be bounded above along the flow.
This seems to eliminate all solutions with $A<1/4$. The case
$A=1/5$ in the examples of N=4 Coulomb branch is indeed known
to correspond to a singular 10$d$ solution with negative tension branes.
The criterion can be also understood
as follows. It selects solutions for which the IR ambiguities noticed
in \cite{gppz2} are absent. The action for a (canonically normalized)
scalar $S=\int e^{4\phi}(\partial\lambda)^2$ predicts an IR
contribution to the condensate
\beq
<O_\lambda >={\delta S\over\delta\lambda}\sim e^{4\phi}\partial\lambda
\sim |y-y_0|^{4A-1} 
\eeq{condens}
for all logarithmic flows. This IR ambiguities diverges when $A<1/4$.
The case $A=1/4$ is borderline. It is possible that, as noticed in 
\cite{gubbb}, only the $A=1/4$ solutions representing vacua have
a {\it physical} interpretation.

\section{The Flow to N=1 SYM} 
We now present a holographic RG flow from N=4 SYM to pure
N=1 SYM in the IR. We find agreement with field theory
expectations: quarks confine, monopoles are screened, and there is
a gau-
gino condensate.

Consider a deformation of N=4 Super Yang-Mills
theory with a supersymmetric mass term for the three
fermions in the chiral N=1 multiplets. In N=1 notations, this is a
mass
term for the three chiral superfields $X_i$
\beq
\int d^2\theta m_{ij} \Tr X_i X_j + {\rm c.c.},
\eeq{m31}
where $m_{ij}$ is a complex, symmetric matrix.

The theory flows in the IR to pure N=1 Yang-Mills, which confines.
To obtain the standard N=1 pure Yang-Mills with fixed scale
$\Lambda$, we need a fine
tuning of the UV parameters, in which the mass $m$ diverges while
the 't Hooft
coupling constant, $x$, goes to zero as an (inverse) logarithm of $m$.
This is outside
the regime of validity of supergravity, which requires a large $x$.
We can think of $m$ as a regulator for
N=1 SYM. When embedded in N=4 SYM, the theory is finite. To get a well
defined N=1 SYM, we remove the cut-off ($m\rightarrow \infty$)
with a fine tuning of the coupling ($x(m)\rightarrow 0$). However,
if we use supergravity, we are in the large $x$ regime. The massive modes
have a mass comparable with the scale of N=1 SYM and
they do not decouple. We can think of this as a theory with an
ultraviolet cut-off.
A good analogy is with lattice gauge theory. $1/m$
corresponds to the lattice spacing. The continuum limit
is obtained with a fine tuning $a\rightarrow 0,g(a)\rightarrow 0$.
However we can study the lattice theory at strong coupling, far from the
continuum limit. A standard computation at strong coupling (by Wilson)
gives the area law. We are just doing analogous computations
with supergravity.
Qualitative features of the theory should hold also at strong
coupling.

The 5-dimensional action for the scalars \cite{grw2}
\beq
L = \sqrt{-g}\left[-{R\over 4}
- {1\over 24}\Tr (U^{-1}\partial U)^2 + V(U)
\right],
\eeq{m32}
is written in terms of a
$27\times 27$ matrix $U$, transforming in the fundamental
representation of
$E_6$ and parametrising the coset $E_6/USp(8)$.
In a unitary gauge, $U$
can be written as $U=e^X, X=\sum_a \lambda_a T_a$,
where $T_a$ are the
generators of $E_6$ that do not belong to $USp(8)$.
This matrix has exactly 42 real independent parameters, which are the
scalars of the supergravity theory. They transform in the following 
$SO(6)$ representations:
$\underline{10_c}$, $\underline{20}$, and $\underline{1_c}$.
The supersymmetric mass term for the chiral multiplets, $m_{ij}$,
transforms as the $\underline{6}$ of $SU(3)\in SO(6)$, and the corresponding
supergravity mode appears in the
decomposition of the $\underline{10}\rightarrow
\underline{1}+\underline{6} +\underline{3}$ of $SU(4)$ under 
$SU(3)\times U(1)$. The term $\underline{1}$ in this 
decomposition corresponds instead to the scalar $\sigma$ dual to the gaugino
condensate in N=1 SYM. 
In principle, a generic non-zero VEV for $m_{ij}$ will
induce non-zero VEVs for other scalars as well, due to the existence
 of linear couplings of $m$ to other fields in the potential.
However, if we further impose $SO(3)$ symmetry
by taking $m_{ij}$ proportional to the identity matrix, a
simple group theory exercise shows that all the remaining fields
can be
consistently set to zero. 
This is true also if we consider a two-parameter Lagrangian depending on both
$m$ and $\sigma$.
This felicitous 
circumstance makes an apparently intractable problem
very simple and exactly solvable.

The actual computation is reported in \cite{gppz3}.
The result for the action for $m$ and $\sigma$ (the reason why we are 
considering both modes will be clear very soon) is
\bea
L &=& \sqrt{-g}\{- {R\over 4}
+ {1\over 2}(\partial m)^2+{1\over 2}(\partial \sigma)^2 + \nonumber\\
& & - \frac{3}{8} [(\cosh{{2m\over \sqrt{3}}})^2
+4\cosh{{2m\over \sqrt{3}}}\cosh{2 \sigma}\nonumber\\
&-&(\cosh{2 \sigma})^2 +4 ]\}.
\eea{m315}
The action has the supersymmetric form ~(\ref{I9})
with $W=-\frac{3}{4} \left(\cosh{\frac{2m}{\sqrt{3}}} 
+ \cosh{2\sigma}\right)$. The first order equations~(\ref{I10}) read
\bea
\dot{\phi}&=& \frac{1}{2}\left(1+ \cosh{\frac{2m}{\sqrt{3}}}\right)\\
\dot{m}&=& - \frac{\sqrt{3}}{2}\sinh{\frac{2m}{\sqrt{3}}},\\
\dot{\sigma}&=& - \frac{3}{2}\sinh{2\sigma}.
\eea{m37}

One interesting feature of the solution is that the equations
can be analytically solved. To the best of our knowledge, there is
only another example of analytically solvable flow, describing the
Coulomb branch of N=4 SYM \cite{freed2}. The solution in our case is:
\bea
\phi(y)&=&\frac{1}{2}\log[2\sinh(y-C_1)] +\nonumber\\
 &+&\frac{1}{6}\log[2\sinh(3y-C_2)],
\label{m316'}\\
m(y)&=&\frac{\sqrt{3}}{2}\log\left[\frac{1+e^{-(y-C_1)}}
{1-e^{-(y-C_1)}}\right], \label{m316''}\\
\sigma(y)&=&\frac{1}{2}\log\left[\frac{1+e^{-(3y-C_2)}}
{1-e^{-(3y-C_2)}}\right].
\eea{m316}

The metric has a singularity at $y=C_1$ with $A=1/2$
\beq
ds^2= dy^2 + |y-C_1| dx^\mu dx_\mu .
\eeq{m39}
Around this point $m$ behaves as
\beq
m \sim -\frac{\sqrt{3}}{2} \log(y-C_1) + \mbox{const}.
\eeq{m310}
Here we assumed that $C_2\le 3 C_1$, so that at the 
point where $m$ is singular, $\sigma$ is still finite. 

Let us notice that 
this intuitive criterion for selecting {\it physical} solutions
is in agreement with the one proposed in \cite{gubbb}, which
exactly selects the solutions with $C_2\le 3C_1$. 
For $C_2> 3C_1$, $\sigma$ diverges first
with a value $A=1/6$. For these and other reasons, we regard
these solutions as {\it unphysical}. 

\subsection{Properties of the solution}
Let us discuss the qualitative properties of the N=1 SYM solution. 

It is easy to see that the solution corresponds
to a true deformation of the gauge theory. Indeed, $m$ approaches the
boundary in the UV ($y \rightarrow \infty$) as $m \sim e^{-y}$, which
is the required behaviour of a deformation (see eq.(\ref{I3})). On the other
hand, $\sigma$ has the UV behaviour appropriate for a condensate
$\sigma \sim e^{-3y}$. Let us stress that this behaviour is enforced
by the requirement of N=1 supersymmetry along the flow. The interpretation
of the solution is therefore the following: upon perturbation with
a mass term for the three chiral fields, the N=4 SYM theory flows
in the IR to pure N=1 SYM in a vacuum with a non-zero gaugino condensate.
The existence of a gaugino condensate is one of the QFT expectations
for N=1 SYM. 

We also expect the gauge theory to exhibit confinement in
the IR. We can easily compute a two-point function for a minimally-coupled
scalar in the background with $\sigma =0$. 
In our example, the Schroedinger potential is 
\beq
V(z)={6\cos (2z)+9\over \sin^2 (2z)}.
\eeq{scpot}

\begin{figure}[h]
\epsfxsize = 2.8in
\epsffile{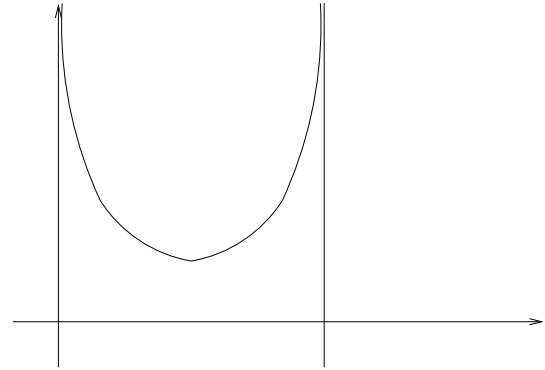}
\caption{The potential for the N=1 SYM flow.}
\end{figure}


It is obvious from 
the figure below that there is mass gap and a discrete spectrum. The AdS boundary
is at $z=0$ and the singularity at $z=\pi /2$.

The two-point function for the massless scalar corresponding to $F^2$
can be explicitly computed \cite{ans}:
\beq
\langle F^2(k) F^2(0)\rangle\sim
k^2 (k^2+4)\mbox{Re\,}\psi(2+ik).
\eeq{2p}
It approaches the conformal expression $k^4\log k$ in
the UV and it is analytic for small $k$, as appropriate for a confining
theory.  
It has poles for $M^2=-k^2=n^2, n=2,3,...$, corresponding to
the $F^2$ glueball states in the spectrum.

Despite the presence of a singularity that invalidates
the supergravity approximation in the IR, the qualitative properties of 
the solution agree with the QFT expectations. There is however a disturbing
point: our solution depends on two independent parameters $C_1$ and $C_2$.
The first one fixes the position of the singularity and it is related
to the magnitude of the mass deformation. The second one is instead
related to the magnitude of the gaugino condensate. We have a 
chirally-symmetric vacuum and, more disturbing, a continuous degeneracy
of vacua with arbitrary small condensate. We certainly expect that the correct
treatment of the singularity and its resolution in string theory fixes
the relation between $C_1$ and $C_2$ in agreement with field theory
expectations. We do not still known how to resolve or deal with the
singularity, therefore we limit ourself to a brief discussion of
the QFT expectations and possible interpretations of the singularity.

\subsection{QFT and string expectations}
Strong coupling QFT results for N=1 SYM have been recently
obtained and differ considerably from
the weak coupling ones \cite{dorey}. At weak coupling, spontaneous
breaking of the chiral symmetry $Z_N$ gives $N$ vacua that only differ
for the phase of the gaugino condensate $<\lambda \lambda >\sim e^{2\pi
ik/N}\Lambda_{{\rm N}=1}^3$. In the large $N$ limit, we obtain a {\it
circle} of vacua. 
The magnitude of the gaugino condensate is fixed in terms of the SYM scale 
$\Lambda_{{\rm N}=1}\sim m e^{-1/3Ng^2}$. 
At strong coupling instead, it was shown in 
\cite{dorey} that there is, at least for $\theta =0$, a distribution of
vacua with condensate $<\lambda \lambda >\sim m^3x^3/j^2,\, j=1,2,...$ with
zero phase. The weakly coupled {\it circle} is lost, the condensate magnitude 
is not fixed and the vacua have an accumulation point at the origin
(zero condensate). 
However, we notice that the structure of vacua found in \cite{dorey}
has many similarities with our supergravity result. As 
independently noticed in \cite{gubbb}, it is tempting to
identify the solution with $C_2=3C_1$ with the $j=1$ vacuum in \cite{dorey}.
The other solutions with $C_2< 3 C_1$ should correspond to
the $j\ne 1$ vacua. To see how the continuum of vacua in supergravity 
is reduced to a discrete numerable set,
we should understand how to include string corrections in our computation.
Notice that the solution with $\sigma =0$, which is not appealing on the
ground of weak coupling intuition, could be nevertheless used as a
(reasonable?) 
approximation for the many vacua with small condensate at strong coupling.

It was also proposed in \cite{gubbb}
to fix the relation between $C_1$ and $C_2$
by considering the finite temperature version of our solution, where
conditions to be imposed at the horizon fix the parameters.
One finds $C_2=3C_1$. This is the only special value for our parameters,
since, exactly for $C_2=3C_1$, the two scalars $m$ and $\sigma$ diverge at
the same point in $y$. 
In SYM the breaking of supersymmetry will select
the vacuum with minimal energy. At weak coupling, where all the vacua
have a condensate with the same magnitude, this procedure should give us
also the value of the N=1 condensate. At strong coupling, with condensates of
almost arbitrary magnitude, this would give information at most about
one particular vacuum ($j=1$?).    

The knowledge of the full 10 dimensional solution would greatly help
in understanding the properties of the RG flow and in studying possible
resolutions of the singularity. It may even happen that the singularity
is an artifact of the dimensional reduction, that disappears in 10$d$.
This happens, for example, in the case of the Coulomb branch of N=4 SYM
\cite{freed2}, where the 10 dimensional background is just a regular
continuous distribution of D3-branes. However, even in this context, 
some other equally nice\footnote{But not satisfying the criterion in
 \cite{gubbb}.}
 5$d$ solutions have a lift to still singular 
10$d$ solutions, representing D3-branes with negative tension. 
The complete ansatz for the 10$d$ lifting
of 5$d$ solutions is known only for a subset of scalars,
the $\underline{20}$,
coming from the KK modes of the internal metric. This is sufficient
to lift all solutions representing the Coulomb branch, but it is not of
help with our solution, where the modes $\underline{10}$ from the
anti-symmetric tensors are excited. 

A ten dimensional interpretation of the N=1 solution in terms of
a background with also D5-branes has been proposed in \cite{p}. We
only notice that the ingredients in this interpretation (D5 and NS-branes)
have been independently suggested in \cite{dorey} on the basis of
the strong coupling QFT analysis. 
 
Finally, we mention that a  mechanism for resolving
singularities in distributions of branes which may help,
after the 10$d$ lifting, has been proposed in \cite{pjp}.

\subsection{The Wilson loop}
A complementary approach for checking confinement is the computation
of a Wilson loop, which should manifest
an area law behaviour.  We need to minimise the action for a string
whose endpoints are constrained on a contour $C$ on the boundary. 
The detailed computation is reported in
\cite{gppz2,gppz3}. In the coordinates used in those papers, the
quark-antiquark energy reads
\beq
E = S/T = \int dx \sqrt{(\partial_x u)^2+f(u)}.
\eeq{m312}
where 
$f(u) = T^2(u)e^{4\phi(u)}$. The phase of the theory can be inferred
by the IR behaviour of this function (see \cite{gppz2} for a review
of the various cases). $T(u)$ is the tension of the 
fundamental (in the case of a quark loop) or of the D1 string (monopole) in five
dimensions. They are in general  non-trivial functions of
the scalar fields. The 5$d$ N=8 gauged supergravity has an $SL(2,Z)$ symmetry
that allows to discriminate electric and magnetic strings. They should
couple to the 5$d$
antisymmetric tensors $B_{\mu\nu}^{I\alpha}$, transforming
in the $(\underline{6},2)$ of $SO(6)\times SL(2,Z)$. The $SO(6)$ index
should account for the orientation of the strings on the five-sphere,
while the $SL(2,Z)$ index should iden-\
tify electric and magnetic quantities.  
On the basis of
naive dimensional reduction from ten dimensions, the tensions can be
read from the coefficients of the kinetic term for the 
antisymmetric tensors. In 10 dimensions, the  tension of the fundamental 
string (or the D1-string) can be read from the NS-NS (or R-R) antisymmetric tensor Lagrangian evaluated in the Einstein frame,
\beq
{1\over T_{F1}^2}H^2_{NS\mbox{-}NS} + {1\over T_{D1}^2}H^2_{R\mbox{-}R}.
\eeq{m11}
A simple Weyl rescaling shows that this property is valid also in the
five-dimensional theory in the Einstein frame.

The kinetic terms for the anti-symmetric ten-\
sors can be computed for
the N=1 SYM solution and behave asymmetrically in the $SL(2,Z)$ indices
\cite{gppz3}. The final result for the   
tensions $T(u)$ of the fundamental strings and of
the D1-strings are, respectively,
\bea
T^2_{F1}&=&4\left(\cosh{\frac{4m}{\sqrt{3}}}+\cosh{\frac{2m}{\sqrt{3}}}\right),\\
T^2_{D1}&=&8 \left(\cosh{\frac{m}{\sqrt{3}}}\right)^2,
\eea{m313}
so that the asymptotic behaviour of the corresponding functions $f(u)$ is
\beq
f_{(q\bar{q})}(u) \sim 1, \,\,\,\ f_{(m\bar{m})}(u) \sim \left|u-C_1 \right|.
\eeq{m314}

It is easy to check that $f_{(q\bar{q})}(u)$ is bounded from below. It
follows that the energy $E\ge cL$, where $L$ is the quark
distance. It can be easily proven 
that it is in fact $E=cL$, implying an area law behaviour
for the Wilson loop, as expected for a confining theory. The IR behaviour
of $f_{(m\bar{m})}(u)$ implies, on the other hand, that monopoles
are screened (see \cite{gppz2} for a review).

There is an apparent contradiction in the previous reasoning.
The 5$d$ dilaton is not running in our solution. If the 10$d$ dilaton were also
constant, the tension for a fundamental string would be proportional to the
tension of a D1-string and the same would be true also after dimensional
reduction to 5 dimensions. The 5$d$ tensions would be then complicated
functions of the scalars, but invariant under $SL(2,Z)$. We instead find
an $SL(2,Z)$ asymmetric result from the N=8 gauged supergravity
evaluated along our solution. A possible way out is to assume that, against
 naive
expectations, the 10$d$ dilaton is not constant. 
Clearly, it also exists the option that the 10$d$ dilaton is constant and that
the argument which determines the 5$d$ tensions via dimensional reduction
is too naive.
However, we are not aware of
any argument that rules out the possibility of a running 10$d$ dilaton. 
Since we are not
expert in reconstructing 10$d$ solutions from 5$d$ ones, we just limit
ourselves to consider this option and perform some very preliminary check
on the equations of motion.  

The 10$d$ dilaton equation of motion is
\beq
\partial^2\phi \sim G_{MNP}G^{MNP}.
\eeq{10$d$}
Therefore, a non-vanishing anti-symmetric tensor is a source for the
dilaton. We can perform a check on our solution at the linearised level.
Consider a generic fluctuation of the  anti-symmetric tensor $B_{ab}=f_I(y)
Y^{I\pm}_{[ab]}$. We refer to \cite{van} for notations and useful
equations. Here $Y^{I\pm}_{[ab]}, a,b=1,...,5$ are harmonic functions
on the five-sphere, transforming in the representation $I$ of $SO(6)$.
They satisfy $\epsilon_{abcde}\partial_cY_{[de]}=\pm 2i(k+2)Y_{[ab]}$,
where $k$ is an integer labelling the harmonic degree. It is then
easy to check that 
\beq
\partial^2\phi\sim {1\over 3}((\partial_yf)^2-(k+2)^2f^2)Y_{[ab]}
Y_{[ab]}.
\eeq{ghjgjhjgh}
In our case ($I=\underline{10}$) $k=1$.  Since we are considering
a deformation of the UV fixed point, $f\sim e^{-x}$,  we see that the dilaton
must run. Notice that instead, considering a different vacuum of the UV theory,
one has $f\sim e^{-3x}$, and the dilaton
remains constant (at least at the first perturbative order).

We still need to check that $Y_{[ab]}Y_{[ab]}\ne 0$. 
There is at least one example where $Y_{[ab]}Y_{[ab]}= 0$: the
$SU(3)\times U(1)$ critical point of the N=8 supergravity,
whose 10$d$ solution is explicitly known \cite{romans}. 
In the product
$\underline{10}\times \underline{10}=\underline{20}+...$, only the indicated
term contains scalar terms ($SO(5)\in SO(6)$ singlets). It is easy to check
that, decomposing $\underline{10}=\underline{1}+\underline{3}+\underline{6}$
under $SU(3)\times U(1)$, the $\underline{1}$ term (related to the 
$SU(3)\times U(1)$ critical point) has vanishing square. The N=1 mass term
$\underline{6}$, however, has non vanishing square.

This argument is certainly not a proof that the 10$d$ dilaton runs.
However, we find this option appealing.
A running of the 10$d$ dilaton would agree with
an interpretation of our solution that includes branes others than the
D3s.
In many respects, the knowledge of the explicit 10$d$ solution would
help us in understanding the system, from the constituent branes to
the fate of the singularity.  
Using a D3-brane probe in the 10$d$ background we could also explicitly 
compute the  running of the gauge coupling along the flow.

\vskip .2in
\noindent
\acknowledgments
The content of this paper has to appear divided in two parts 
in the proceedings for 
the TMR conference in Paris, September 99. For sake of economy
and not to bother the potential readers,
we just unified the two parts in the version for the archives. 
We would like to thank our collaborators L. Girardello and M. Porrati,
with whom most of the results reported here were obtained. We also
thank D. Anselmi for useful discussions and collaboration at various
stages. We also thank N. Dorey, S. S. Gubser, S. P. Kumar, N. Warner, C. Pilch
and E. Witten for useful discussions and criticisms. 
A. Z. is partially supported by INFN, and
by the European Commission TMR program ERBFMRX-CT96-0045,
wherein he is associated to the University of Torino. M. P is partially
supported by INFN, MURST, the European Commission TMR program
ERBFMRX-CT96-0045, 
wherein she is associated to Imperial College, London, and the PPARC
SPG grant PPA/G/S/1998/00613.


\begin{thebibliography}{6666666666}
\bibitem{gppz1} L. Girardello, M. Petrini, M. Porrati and A. Zaffaroni,
               \jhep{12}{1998}{022}, hep-th/9810126.

\bibitem{dz} J. Distler and F. Zamora, \atmp{2}{1998}{1405},
             hep-th/9810206.

\bibitem{pilch} A. Khavaev, K. Pilch and N. P. Warner, New Vacua of
              Gauged N=8 Supergravity, hep-th/9812035.

\bibitem{lust} A. Karch, D. L{\"u}st and A. Miemiec, \plb{454}{1999}{265}, 
hep-th/9810254.

\bibitem{ks} A. Kehagias and K. Sfetsos, \plb{454}{1999}{270},
           hep-th/9902125; S. S. Gubser, Dilaton-driven confinement,
hep-th/9902155; N. R. Constable and R. C. Myers, \jhep{11}{1999}{020}, 
hep-th/9905081.

\bibitem{gppz2} L. Girardello, M. Petrini, M. Porrati and A. Zaffaroni,
              \jhep{05}{1999}{026}, hep-th/9903026.

\bibitem{Od} S. Nojiri and S. D. Odintsov, \plb{449}{1999}{39},
        hep-th/9812017; \plb{458}{1999}{226}, hep-th/9904036.

\bibitem{freed1} D. Z. Freedman, S. S. Gubser, K. Pilch and N. P. Warner,
        Renormalization Group Flows from Holography--Supersymmetry and a
       c-Theorem, hep-th/9904017.

\bibitem{freed2} D. Z. Freedman, S. S. Gubser, K. Pilch and N. P. Warner,
         Continuous distributions of D3-branes and gauged supergravity,
          hep-th/9906194; 

\bibitem{brr}A. Brandhuber and K. Sfetsos, Wilson loops 
          from multicentre and rotating branes, mass gaps and phase structure 
          in gauge theories, hep-th/9906201; I. Chepelev and R. Roiban,
          \plb{462}{1999}{74}, hep-th/9906224. I. Bakas and K. Sfetsos,
States and Curves of Five-Dimensional Gauged Supergravity, hep-th/9909041.

\bibitem{gppz3} L. Girardello, M. Petrini, M. Porrati and A. Zaffaroni,
               The Supergravity Dual of N=1 Super Yang-Mills Theory
hep-th/9909047.

\bibitem{ps} M. Porrati and A. Starinets, Phys. Lett. B454 (1999) 77,
hep-th/9903085.

\bibitem{bk} V. Balasubramanian and P. Kraus, \prl{83}{1999}{3605},
  hep-th/9903190.

\bibitem{verlinde} J. de Boer, E. Verlinde and H. Verlinde, hep-th/9912012.

\bibitem{bala} V. Balasubramanian, P. Kraus, A. Lawrence and  S. Trivedi,
         \prd{59}{1999}{104021}, hep-th/9808017.

\bibitem{kw2} I. R. Klebanov and E. Witten, \npb{556}{1999}{89},
  hep-th/9905104.

\bibitem{ans} D. Anselmi, L. Girardello, M. Porrati and A. Zaffaroni,
A Note on the Holographic Beta and C Functions, hep-th/0002066.

\bibitem{gubbb} S. Gubser, Santa Barbara ITP Conference on 
New Dimensions in Field Theory and String Theory 
November 17-20, 1999 http://online.itp.ucsb.edu\\ /online/susy$\_$c99/gubser;
and now appeared as hep-th/0002160.

\bibitem{townsk} K. Skenderis and P. K. Townsend, \plb{468}{1999}{46},
hep-th/9909070.



\bibitem{cvet} M. Cvetic, S. S. Gubser, H. Lu and C. N. Pope, 
Symmetric Potentials of Gauged Supergravities in Diverse Dimensions and 
Coulomb Branch of Gauge Theories, hep-th/9909121.

\bibitem{pjp} C. V. Johnson, A. W. Peet and J. Polchinski, Gauge Theory and 
the Excision of Repulson Singularities, hep-th/9911161.

\bibitem{w2} E. Witten, \atmp{2}{1998}{505}, hep-th/9803131.

\bibitem{rs} L. Randall and R. Sundrum, \prl{83}{1999}{4690}.

\bibitem{malda} J. Maldacena, \atmp{2}{1998}{231}, hep-th/9711200.

\bibitem{PP} A. Peet and J. Polchinski, \prd{59}{1999}{065011},
hep-th/9809022.

\bibitem{kw1}I. R. Klebanov and E. Witten, \npb{536}{1998}{199}, 
hep-th/9807080.

\bibitem{g} S. S. Gubser, \prd{59}{1999}{025006},  hep-th/9807164.

\bibitem{gk}S. S. Gubser and I. R. Klebanov, \prd{58}{1998}{125025},
hep-th/9808075.
\bibitem{sergio} A. Ceresole, G. Dall'Agata, R. D'Auria and S. Ferrara,
\prd{61}{2000}{066001}, hep-th/9905226.
\bibitem{warn} M. G{\"u}naydin, L.J. Romans and N.P. Warner, \plb{154}{1985}
{268}; M. Pernici, K. Pilch and P. van Nieuwenhuizen, 
\npb{259}{1985}{460}.


\bibitem{afgj1} D. Anselmi, D.Z.Freedman, M.T. Grisaru and A.A. Johansen,
\npb{526}{1998}{543}, hep-th/9708042.

\bibitem{nsvz} V. Novikov, M.A. Shifman, A.I. Vainshtein and
V. Zakharov, \npb{229}{1983}{381}.

\bibitem{hs} M. Henningson and K. Skenderis, \jhep{07}{1998}{023},
hep-th/9806087.

\bibitem{grw2} M. G{\"u}naydin, L.J. Romans and N.P. Warner, \npb{272}
{1986}{598}.

\bibitem{non} D. Anselmi, L. Girardello, M. Petrini and A. Zaffaroni,
unpublished.

\bibitem{afgj2} D. Anselmi, J. Erlich, D.Z.Freedman, and A.A. Johansen,
\prd{57}{1998}{7570}, hep-th/9711035.

\bibitem{ans2} D. Anselmi, Towards the classification of conformal field theories in arbitrary even dimension, hep-th/9908014.

\bibitem{w1} E. Witten, \atmp{2}{1998}{253}, hep-th/9802150.

\bibitem{fz} S. Ferrara and A. Zaffaroni, \plb{431}{1998}{42}.

\bibitem{gkp} S.S. Gubser, I.R. Klebanov and A.M. Polyakov, \plb{428}
               {1998}{105}, hep-th/9802109.

\bibitem{oz} C. Csaki, H. Ooguri, Y. Oz and J. Terning, 
\jhep{9901}{1999}{017}, hep-th/9806021.


\bibitem{p} J. Polchinski, Caltech Conference,\\
 String at the millenium,
12-15 January 2000,
 Caltech,
http://quark.theory.\\
caltech.edu/people/rahmfeld/Polchin-\
ski/fs1.html.
\bibitem{dorey} N. Dorey and S. P. Kumar, Softly-Broken N = 4 
Supersymmetry in the Large-N Limit, hep-th/0001103.
\bibitem{van} H. J. Kim, L. J. Romans and P. van Nieuwenhuizen, \npb{242}
{1984}{377}.
\bibitem{romans} L. J. Romans, \plb{153}{1985}{392}.
\end{thebibliography}
\end{document}